\documentstyle[11pt,newpasp,twoside]{article}
\newcommand{\be}{\begin{equation}}
\newcommand{\ee}{\end{equation}}
\newcommand{\ba}{\begin{eqnarray}}
\newcommand{\ea}{\end{eqnarray}}
\newcommand{\siml}{\lower4pt \hbox{$\buildrel < \over \sim$}}
\newcommand{\simg}{\lower4pt \hbox{$\buildrel > \over \sim$}}
\newcommand{\Mesz}{M\'esz\'aros}
\def\ref{\par \hangindent 0.3 cm\noindent}

\def\edcomment#1{\iffalse\marginpar{\raggedright\sl#1\/}\else\relax\fi}
\reversemarginpar

\begin{document}
\title{Some recent developments in gamma-ray burst afterglow and
prompt emission models}\footnote{Wang is the presenter of
the talk at the conference}
 \author{Bing Zhang, Peter M\'esz\'aros \& Junfeng Wang}
\affil{Department of Astronomy \& Astrophysics, Pennsylvania State
University} 

\begin{abstract}
Extensive observational campaigns of afterglow hunting have greatly
enriched our understanding of the gamma-ray burst (GRB) phenomenon.
Efforts have been made recently to explore some afterglow properties
or signatures that will be tested by the on-going or the future
observational campaigns yet come. These include the properties of GRB
early afterglows in the temporal domain; the GeV-TeV afterglow
signatures in the spectral domain; as well as a global view about the
GRB universal structured jet configuration. These recent efforts are
reviewed. Within the standard cosmological fireball model, the very
model(s) responsible for the GRB prompt emission is (are) not
identified. These models are critically reviewed and confronted with
the current data. 
\end{abstract}

\section{Introduction}

Gamma-ray bursts (GRBs) belong to one of those classes of mysterious
objects whose nature took great efforts to identify. For years after
their discovery, our knowledge about the objects had been limited to
the ``gamma-ray'' in the spectral domain, and the ``burst'' in the
temporal domain. The revolutionary discovery and the extensive hunting
and monitoring of the broadband afterglows for dozens of GRBs greatly
extended our knowledge about the events both in the spectral domain
(from radio to X-rays) and in the temporal domain (from minutes to
years). We now have some good knowledge about at least one subset of
GRBs, i.e. the so-called long GRBs whose durations are longer than 2
seconds. These include that they are originated from cosmological
distances, likely associated with the deaths of massive stars; that
the afterglow is emission from the external forward shock when the
fireball impacts the ambient interstellar medium (ISM); and that the
fireballs are likely collimated at least for some GRBs (e.g., Piran
1999; van Paradjis, Kouveliotou, \& Wijers 2000; M\'esz\'aros
2002). It is expected that some new progress will be made in the GRB
study in the coming years, thanks to the observational advances led by
some future space and ground-based multi-wavelength telescopes,
especially by the two NASA missions, Swift and GLAST. Among others,
the very early afterglows including the bridge between the GRB prompt
phase and the afterglow phase will be carefully studied; the
broad-band afterglow campaign will be extended to high energy
(GeV-TeV) regimes; and a large sample of dataset combining the
redshifts and the GRB prompt emission \& afterglow properties will be
available, leading to a global view of the GRB jet configuration and
the identification of the GRB prompt emission site(s) and
mechanism(s). Below we will review some recent theoretical efforts in
anticipation of these upcoming observational breakthroughs.

\section{Early afterglows}

GRBs involve extremely relativistic bulk motion of the fireballs. At
the time when most of the current afterglow observations are made
(typically hours after the burst trigger), a fireball has already
decelerated substantially to mildly relativistic (e.g. $\Gamma(t) \sim
20 (E_{52}/n)^{1/8} [t_h/(1+z)]^{-3/8}$, where $E_{52}$ is the
isotropic fireball energy in unit of $10^{52}$ ergs, $n$ is the ISM
density, $z$ is the redshift, and $t_h$ is the epoch of observation in
the observer frame in unit of hour), and the erratic central
engine activity has usually ceased. Catching the detailed information
in the very early epoch of the afterglow emission is of great interest 
in catching precious information about the fireball relativistic
motion and the central engine behavior.

\subsection{Early post-injection signatures} 

The observer's time at the on-set of the afterglow emission is
approximately $t_{dec} \sim 2.4~{\rm s}~ (E_{52}/n)^{1/3}$
$(\Gamma_{0}/300)^{-8/3} (1+z)$, where $\Gamma_0$ is the ``initial''
bulk Lorentz factor of the fireball at the deceleration radius, which
is scaled with a typical value, 300. This timescale is largely
determined by the fireball intrinsic parameters $E_{52}$ and
$\Gamma_0$, as well as the environmental parameter $n$. The time scale
for the central engine activity, $t_{eng}$ on the other hand, is
completely defined by the life time of the central engine (e.g. the
life time of the black hole - torus system; the collapsar envelope
fallback time scale; or the life time of the temporal strong toroidal
magnetic field for the magnetar central engine case). This is usually
manifested as the duration of the GRB (but in principle could be
longer), which is found to vary substantially among bursts. These two
time scales are independent, and as long as $t_{eng} > t_{dec}$,
post-energy-injection into the fireball in the afterglow phase is
inevitable. 

Whether the post-injection behavior is noticeable depends on the
comparison between the initial energy, $E_{imp}$, already entrained in
the fireball and the additional energy input, $E_{inj}$. Only if the
latter is comparable to or larger than the former, could the injection
process leaves noticeable signatures in the afterglow lightcurves
(Zhang \& M\'esz\'aros 2002a). Also the central engine activity is
expected to diminish at later times. Therefore the early afterglow
phase is the most likely epoch to study the injection signatures.

The injection signatures depend on the nature of the injected energy
flow. 1. If the GRB wind is mainly dominated by a cold, magnetic
dominated component, as some GRB central engine models expect, the
injection wind is likely to give an achromatic, gradual bumping
signature in the broad-band afterglow lightcurves (Fig.1 of Zhang \&
M\'esz\'aros 2002a). One example of such a
Poynting-dominated-injection is that from a strongly magnetized
millisecond pulsar (Dai \& Lu 1998; Zhang \& M\'esz\'aros 2001a),
which presents a well-defined, easy-modeling injection signature
(Zhang \& M\'esz\'aros 2002a). A detection of such a signature can
give a possible identification of the GRB central engine otherwise
poorly known. Whether a strong reverse shock is developed in such an
injection case is unclear, and the contribution from the reverse shock
is neglected in the current modeling. 2. In the conventional ``hot''
fireball scenario, the injected energy is typically in the form of the
kinetic energy of the expanding shell.  In a simple case, a fireball
is composed of multi-minishells with increasing Lorentz factors at
later times. Such a scenario is known as the ``refreshed'' shock
scenario (Rees \& M\'esz\'aros 1998; Sari \& M\'esz\'aros 2000), and
there is already a claim that such a scenario may be able to interpret
the peculiar flat decaying lightcurves of GRB 010222 (Bj\"ornsson et
al. 2002). In more complicated scenarios, an injection process is more
appropriately viewed as the interactions among three shells (ISM,
initial shell, and the injective shell). The hydrodynamics of such a
process is analyzed in great detail in Zhang
\& M\'esz\'aros (2002a). Because multi-shocks are involved, the injection
signatures for a violent collision are complicated (Fig.5 in Zhang \&
M\'esz\'aros 2002a). The recent peculiar wiggling R-band lightcurve of
GRB 021004 may be modeled as such violent injection events, although
other interpretations (e.g. Lazzati et al. 2002) may be also
reasonable.

Other mechanisms to interpret lightcurve bumps include supernova
contribution (e.g. Bloom et al. 1999), dust echo (Esin \& Blandford
2000), collisions of the fireball with the ambient density bumps (Dai
\& Lu  2002a), and the gravitational microlensing effect (Garnavich,
Loeb \& Stanek 2000). The main differentiable character of an
injection signature is that the afterglow level is systematically
raised after the injection (Zhang \& M\'esz\'aros 2002a).

\subsection{Reverse shock emission}

During the initial interaction of the fireball with the ISM, a reverse
shock will cross the fireball shell and emit a comparable amount of
energy as in the forward shell (M\'esz\'aros \& Rees 1997). After the
shell crossing, the shocked fireball cools adiabatically, leading to a
fading emission component. The emission is peaked in the UV-optical
band, and it has been claimed that the optical flash detected in GRB
990123 is due to such an emission component (M\'esz\'aros \& Rees
1999; Sari \& Piran 1999). Due to complicated regimes involved
(Kobayashi \& Sari 2000; Kobayashi 2000), a bright optical flash is
not always expected to be detectable. In any case, when early
afterglows are caught, the emission signatures dominated by the
reverse shock should be attainable.

Thanks to the prompt localization by HETE II, lately the R-band
afterglow of GRB 021004 was caught $\sim$ 9 minutes after the burst,
marking the earliest record of catching GRB afterglows and the best
monitored afterglow so far (see extensive 
observational reports in the GRB Coordinates Network,
http://gcn.gsfc.nasa.gov/gcn/gcn3\_archive.html). An apparent
re-brightening in the R-band lightcurve is observed at around 0.1
day. Among other interpretations (e.g. Lazzati et al. 2002), Kobayashi
\& Zhang (2002) show that this re-brightening is well interpreted
within the framework of the standard fireball model in which both the
emissions from the forward shock and from the reverse shock are
considered. The low-frequency (e.g. optical, IR \& radio) lightcurve
from the forward shock is expected to rise initially and to decay
after the typical synchrotron frequency $\nu_m$ crosses the
observational band (Fig.2b of Sari, Piran \& Narayan 1998). Although
such a behavior has been observed in the radio band, previous optical
afterglow observations were only made at too late a time to catch the
rising part of the lightcurve. Kobayashi \& Zhang (2002) argue that
such a phase was caught for GRB 021004, together with a rapid decaying
($F_\nu \propto t^{-2}$) emission component from the reverse shock.
The superposition of both the forward shock and the reverse shock
emissions can well fit for the 0.1-day re-brightening feature of the
GRB 021004 R-band early afterglow. For the example case presented in
Kobayashi \& Zhang (2002), the standard shock parameters as inferred
from other GRBs (Panaitescu \& Kumar 2001, 2002) are adopted, and the
model can also interpret the X-ray and radio afterglow data.  This
hints that such a re-brightening signature may be a common feature in
early afterglows, which could be readily tested in the near future by
the UVOT telescope on board Swift mission.

\section{High energy afterglows}

The term ``high energy afterglow'' is defined with respect to the
conventional low-energy (X-ray, optical, radio) afterglows which are
believed to be originated from the synchrotron emission of the
electrons accelerated in the forward shocks as the fireball
decelerates in the ISM. The origins of the possible high energy
afterglows are well justified. There are several high energy spectral
components in the standard shock scenario which are not important in
the low energy band. These include the synchrotron self-inverse
Compton (IC) emission of the electrons, and the baryon-related
emission components such as the proton synchrotron emission,
synchrotron emission from the secondary muons or direct $\pi^0$ decay
from the photomeson interactions. Zhang \& M\'esz\'aros (2001b)
presented a coherent study of the high energy spectral components with
respect to the conventional synchrotron component, and discussed the
shock parameter regimes in which various components dominate. Because
baryons are redundant emitters compared with electrons, the
baryon-related emission components can overwhelm the electron
synchrotron emission only when the electron equipartition factor
$\epsilon_e$ is very small while the magnetic field is near
equilibrium ($\epsilon_B \sim 1$). On the other hand, there is a large
parameter regime in the $\epsilon_e-\epsilon_B$ space that the IC
component is important (regime II of Fig.1 in Zhang \& M\'esz\'aros
2001b, typically $\epsilon_e/\epsilon_B>1$). It turns out that the
current broadband afterglow fits result in typical $\epsilon_e$ and
$\epsilon_B$ parameters which lie right in this IC-dominated regime
(Panaitescu \& Kumar 2001).

A direct consequence for these ``typical'' (regime II) GRBs is that
the IC-origin high energy afterglows should be detectable. The IC
emission forms a separate four-segment broken power law spectral
component at the high energy end of the conventional synchrotron
component (e.g. Sari \& Esin 2000). As the blast wave decelerates, the
two-hump spectrum will evolve towards down-left. Fixing a certain
band, the observer will initially receive power-law-decaying
synchrotron emission until, at a critical time $t_{_{\rm IC}}$, the IC
component overtakes the synchrotron component. For the electron
power-law index $p=2.2$, this critical time is (Zhang \& M\'esz\'aros
2001b) 
\begin{equation}
t_{_{\rm IC}}=3.4~{\rm days}~(\epsilon_e/0.5)^{0.89} (\epsilon_B /
0.01)^{0.08} E_{52}^{-0.06} n^{-0.66} (1+z)^{-0.36} \nu_{18}^{-0.68}
\end{equation}
After $t_{_{\rm IC}}$, the lightcurve then climbs up until reaching
the peak flux of the IC component, and declines later after
$\nu_{m,{\rm IC}}$ crosses the band. This results in a bumping
signature in the fixed band lightcurve. Equation (1) shows that in
order to have a noticeable signature at an early time, the frequency
has to be high enough (X-ray upwards), and a dense environment tends
to ease the IC-dominant condition. Late X-ray bump in the GRB 000926
has been attributed to the IC high energy afterglow (Harrison et
al. 2001).

At higher frequencies, the IC-dominated era as well as the IC peak
time move to earlier epochs. For the nominal GLAST band (0.4-200 GeV),
the afterglow is dominated by the IC component as soon as the
afterglow starts, and the IC peak sweeps the band in about an
hour. Given the GLAST sensitivity, a typical regime II GRB at $z=1$
could be detected by GLAST from minutes to about a day after the
trigger, leading to an extended high energy afterglow (Fig.4 of Zhang
\& M\'esz\'aros 2001b). According to this scenario, the long term GeV
emission detected by EGRET in GRB 940217 (Hurley et al. 1994) may have
been due to such an IC component for a nearby GRB (see also
M\'esz\'aros \& Rees 1994; Dermer, Chiang \& Mitman 2000).

Besides this standard model, an extended GeV afterglow may be also
interpreted by some alternative mechanisms. One possibility is the
secondary IC scattering off the microwave background photons by the
pairs produced by the TeV gamma-rays generated in the GRB internal
shocks through interacting with the IR background photons (Dai \& Lu
2002b), if the intergalactic magnetic field is not strong enough for
prominent lepton reflection to occur. Another possibility is the
direct interaction of the GRB fireball with the hot ambient pulsar
wind nebula (Wang, Dai \& Lu 2002; Granot \& Guetta 2002). Future
detailed data from GLAST and other space or ground-based high energy
telescopes will provide clues to differentiate among these scenarios.

\section{A quasi-universal structured jet model}

Achromatic steepening of the afterglow lightcurves infer that most
long GRBs, if not all, are originated from collimated jets. Frail et
al. (2001) collected a sample of GRBs with redshift and jet break
measurements before Oct. 2000, and found the remarkable fact that the
product of the ``isotropic'' gamma-ray energy and the square of the
inferred jet angle is a standard value, i.e., $E_\gamma=E_{iso}
\theta_j^2 \sim {\rm const}$. The $\theta_j$ angle is obtained from
the jet breaking time by assuming the simplest uniform sharp-edge jet
configuration (Rhoads 1999; Sari, Piran \& Halpern 1999; Panaitescu \&
M\'esz\'aros 1999; Moderski et al. 2000; Huang et al. 2000; Granot et
al. 2001) and by assuming a roughly constant ambient ISM density ($n
\sim 1~{\rm cm}^{-3}$). Although $n$ could in principle vary from
burst to burst, that most of the later-observed GRBs seem to fall into
the Frail et al. (2001)'s empirical law suggests that the collimated
jets are indeed the promising interpretation of the lightcurve breaks,
and that $n$ is not fluctuating significantly.  Within the
conventional uniform jet model, the jet aperture solid angle is
proportional to $\theta_j^2$, so that the finding of Frail et
al. (2001) suggests that long GRBs have a quasi-standard energy
reservoir, although such a standard energy budget is distributed with
different concentrations among different bursts.

The same data may be interpreted within another, probably more
elegant, picture. Stimulated by Frail et al (2001)'s result, Rossi,
Lazzati \& Rees (2002a) and Zhang \& M\'esz\'aros (2002b) have
independently shown that the wide variety of burst phenomenology could
be attributed to a standard non-uniform jet configuration being viewed
from different orientations. The jet is structured such that the
closer to the jet axis, the more energy is concentrated and the higher
the bulk Lorentz factor is achieved. The jet break angles as inferred
from the observational break times are a manifest of the observer's
viewing angle rather than the real jet opening angle. The shape of the
jet structure is modeled as a power-law, i.e., $\epsilon(\theta)
\propto
\theta^{-k}$, and for $k \sim 2$, such an interpretation is almost
non-differentiable from the conventional uniform jet model. Zhang \&
M\'esz\'aros (2002b) further proposed that the observed jet breaks can
be interpreted within such a scenario for a much broader categories of
the energy distribution functions (e.g. no longer a power law, or even
not a simple analytical form) of the jets. The lightcurve predictions
for these configurations as well as for all the $k\neq 2$ power-law
configurations are not solely equivalent to the predictions from the
conventional jet model (R. Sari, personal communication), but the
conventional model is not fully satisfactory anyway (e.g. a large
dispersion of the electron power law indices $p$ as inferred from the
broadband afterglow fits, while computer simulations on the shock
acceleration indicate that $p$ is likely universal). A closer study on
the structured jet model both analytically and numerically (although
may not be easy) is therefore desirable.

Although the conventional jet model is not easily ruled out, there are
already several arguments in favor of the universal structured jet
model.  1. To interpret the spectral-lag vs. luminosity correlations
found in the GRB prompt emission data (Norris et al. 2000), one
natural way is to incorporate a structured jet configuration and the
line-of-sight effect (Salmonson 2001; Salmonson \& Galama 2002; Norris
2002). A coherent picture is achievable for both the prompt emission
and the afterglow within the framework of the universal jet
model. 2. Within the top candidate of the progenitor models, i.e., the
collapsar model, a structured jet configuration is naturally expected
from numerical modeling (Woosley et al. 2002; Zhang et al. 2002). 3. A
prediction about GRB luminosity function is available for the
universal jet model (but not for the conventional jet model), which is
a power law with index around 2 (Rossi et al. 2002a; Zhang \&
M\'esz\'aros 2002b). Such a prediction is consistent with the GRB
luminosity function derived from the prompt emission data (Schaefer et
al. 2001; Schmidt 2001) or even with the one measured directly from
those bursts with luminosity information (Perna, Sari \& Frail 2002).

There are several ways to differentiate the universal jet model from
the conventional jet model. 1. Besides the well-defined luminosity
function testable in the Swift era (which can add proofs or
constraints on the universal jet model but not on the conventional jet
model due to the lack of the prediction power of the latter), the
GRB-to-orphan afterglow rate are different between different models
(e.g. Totani \& Panaitescu 2002). Future detections of the orphan
afterglows as well as their statistics may shed light onto the issue,
although the orphan afterglows arising from the dirty fireballs may
complicate the clean picture (Huang, Dai \& Lu 2002). 2. The different
afterglow polarization predictions for both models may be helpful to
tell the difference (Rossi et al. 2002b). 3. Future gravitational
radiation polarization data may shed light on the correctness of the
universal jet scenario (Kobayashi \& M\'esz\'aros 2002).

\section{Prompt emission models}

Despite of the extensive observational campaigns and theoretical
modeling of the afterglows, the nature of the GRB prompt emission is
not fully understood. The location and mechanism of the prompt
emission or even the energy context of the fireball are not
unambiguously identified, leading to a variety of the cosmological
fireball model variants.

Most generically, a fireball consists of a hot component with
luminosity $L_h$ (which is essentially the conventional fireball
initially composed of in-equilibrium photons and pairs, and which is
accelerated and coasted until the energy is stored in the form of the
kinetic energy) and a cold component with luminosity $L_c$ (which is
essentially of the form of a Poynting flux, analogous to the case of
a pulsar wind). A parameter, $\sigma \equiv L_c/L_h$, can be defined
to categorize the fireball. Based on the proposed emission site of the 
GRB prompt emission, the GRB models can be also divided into the
external models (where the prompt emission occurs at the deceleration
radius, i.e., $r=r_{dec}$), internal models (where the prompt emission
occurs before deceleration but beyond the photosphere, i.e. $r_{ph}<
r< r_{dec}$), and the innermost models (where the prompt emission
occurs right above the photosphere, i.e. $r \simg r_{ph}$). A unified
picture of these model variants is analyzed in Zhang \& M\'esz\'aros
(2002c).

It is not a easy task to identify the very mechanism(s) responsible
for the observed GRB prompt emission. In any case, for those
well-studied models, clear predictions are available which could be
directly compared against the current and future data. Besides the
complex lightcurve information which requires erratic central engine
behaviors or a clumpy ambient environment, the spectral information,
especially the information about the spectral break, $E_p$, can
provide some useful clues. Two interesting issues include (1) the
distribution of the measured $E_p$ within different bursts or among
various epochs of one single burst (which is found to be narrowly
distributed for the bright BATSE bursts, Preece et al. 2000), and (2)
the possible correlation of $E_p$ with other measurable parameters
(e.g. a positive correlation between $E_p$ and luminosity $L$, Amati
et al. 2002; Lloyd-Ronning \& Ramirez-Ruiz 2002). Assuming synchrotron
radiation as the mechanism, Zhang \& M\'esz\'aros (2002c) have
compiled all the predictions of $E_p$ as functions of various
measurable parameters as well as some unknown parameters (their Table
1). Adopting some distribution functions of the parameters (either
from the observations or from the assumptions), they modeled the
narrowness of the $E_p$ distributions for various models. The results
are compared against the current data. Although any unambiguous clue
for the identification of the ``right'' model is not yet available,
some important constraints have been posed on various models.

1. About the width of $E_p$ distribution: None of the current GRB
models can reproduce the narrow $E_p$ distribution found by Preece et
al. (2000), including the well-discussed internal shock and the external
shock models. A narrow $E_p$ distribution favors those models whose
$E_p$ prediction relies on less free parameters with low power
indices. A high $\sigma$ internal model or a pair-dominated internal
model are good in this respect due to their (essentially) constant
$E'_p$ value in the comoving frame. However, these models are less
studied compared to the standard optically-thin shock scenarios. There
are two issues that may bring data closer to the predictions in the
shock models.  First, the growing population of the so-called X-ray
flashes (XRFs, Heise et al. 2001; Kippen et al. 2001) may broaden the
real $E_p$ distribution. Second, there may be some intrinsic
conspiracy among some parameters in the shock models which tend to
narrow the final $E_p$ distribution. In any case, the internal
scenarios are generally favored against the external scenarios, and
the IC-origin GRB models are less favored than the synchrotron models
since they tend to amplify the $E_p$ scatters. A recent
``supercritical pile'' model invoking resonant baryonic pair
production instability (Kazanas et al. 2002) seems to be favorable in
interpreting the narrow $E_p$ distribution. It is of interest to
investigate the compatibility of this model with other observational
aspects. 

2. About the positive $E_p-L$ correlation: Whether a model is
compatible with the positive $E_p-L$ correlation (Amati et al. 2002;
Lloyd-Ronning 2002) depend on the unknown $L-\Gamma$ correlation,
where $\Gamma$ is the bulk Lorentz factor. For the most
straightforward scenario in which $\Gamma$ is positively correlated
with $L$, the internal shock model is not favored, essentially because
the model predicts a lower $E_p$ for a larger $\Gamma$ since a larger
$\Gamma$ corresponds a larger shock dissipation radius where the
magnetic field is lower. The internal high-$\sigma$ model or the
internal pair-rich model are however compatible with the data. In
order to make the internal shock model remain a promising candidate,
essentially no correlation between $L$ and $\Gamma$ is required.

3. About the nature of X-ray flashes (XRF): The XRFs (Heise et
al. 2001; Kippen et al. 2001) and GRBs are likely different
appearances of a same type of event. In various models, the $E_p$
scatter is caused by the contribution of the scatters of many
independent parameters. Thus it is not straightforward to regard the
scatter of one particular parameter (e.g., Lorentz factor or redshift)
as the reason for the XRF/GRB discrimination. Nonetheless, a direct
reasoning from the ``Lorentz boost'' argument is that XRFs are dirty
fireballs while GRBs are clean fireballs (e.g. Dermer et
al. 1999). This interpretation, together the standard structured jet
configuration as discussed in Section 4, suggests that XRFs are those
GRBs viewed at large viewing angles (e.g. Woosley et al. 2002). A
premise of such an interpretation, however, is that prompt emission is
not from internal shocks because the internal shock model expects high
$E_p$'s for dirty fireballs. An alternative interpretation of the XRFs
and X-ray rich GRBs is in terms of the standard shock model with the
dominant emission components from either the baryonic or the pair
photospheres (M\'esz\'aros et al. 2002). Spectral and redshift
information for XRFs would be essential to testify these
interpretations.

\section{Conclusions}

A new epoch for the GRB study is coming within the next several years,
especially following the launches of Swift and GLAST, and the
utilization of some other broad-band advanced facilities. It is
optimistic to expect the following breakthroughs in the coming years:

\ref $\bullet$ Early afterglows will be carefully studied. The missing 
link between the prompt emission and the afterglow will be identified,
including the detailed reverse shock information. The information from
the central engine may be retrievable through well-modeled injection
signatures.
\ref $\bullet$ High energy afterglows will be monitored and studied in 
conjunction with the low energy afterglows, which will bring invaluable
information about the unknown shock physics and the GRB environments.
\ref $\bullet$ The GRB jet configuration will be identified. We expect 
the universal structured jet model will be validated by future data.
\ref $\bullet$ With accumulation of a large sample of the redshift and 
spectral information for GRBs/XRFs in the Swift era, the right
emission site(s) and mechanism(s) for the prompt emission may be
identified (or at least strongly constrained). The fireball content
and the nature of the relativistic wind from the central engine may be
also understood.

\acknowledgements {We thank Shiho Kobayashi for stimulating
collaborations. This work is supported by NASA grants NAG5-9192 and
NAG5-9153. }

\end{document}